\documentclass[showpacs,byrevtex,aps,preprint]{revtex4}
\usepackage{epsfig,psfig}

\begin{document}


\title{Critical behavior of an epidemic model of drug resistant diseases}

\author{C. R. da Silva$^1$, U. L. Fulco$^2$, M. L. Lyra$^1$, and
G. M. Viswanathan$^1$}

\affiliation{Departamento de F\'{\i}sica, Universidade
Federal de Alagoas, Macei\'{o}--AL, 57072-970, Brazil}

\affiliation{$^1$Departamento de F\'{\i}sica, Universidade
Federal de Alagoas, Macei\'{o}--AL, 57072-970, Brazil}

\affiliation{$^2$Departamento de F'{\i}sica, Universidade Federal
do Piau\'{\i}, Teresina--PI, 64049-650, Brazil}

\begin{abstract}
In this work, we study the critical behavior of an epidemic
propagation model that considers individuals that can develop drug
resistance.  In our lattice model, each site can be found in one
of four states: empty, healthy, normally infected (not drug
resistant) and strain infected (drug resistant) states.  The most
relevant parameters in our model are related to the mortality,
cure and mutation rates. This model presents two distinct
stationary active phases: a phase with co-existing normal and drug
resistant infected individuals and an intermediate active phase
with only drug resistant individuals.  We employ a finite-size
scaling analysis to compute the critical points the critical
exponents ratio $\beta/\nu$ governing the phase-transitions
between these active states and the absorbing inactive state. Our
results are consistent with the hypothesis that these transitions
belong to the directed percolation universality class.
\end{abstract}
\date{\today} 


\maketitle



\section{Introduction}
The spread of epidemic processes in a population has been
profusely studied by biologically oriented mathematicians mainly
by the use of a set of partial differential equations governing
the time evolution of global variables such as the density of
infected, refractory and healthy individuals~\cite{book}. At high
dimensions where local density fluctuations are negligible, this
mean-field like approach gives the correct overall behavior of
these systems, according to renormalization group arguments. At
low dimensions, density fluctuations become relevant and
microscopic approaches have to be considered. In general, most
models of epidemic propagation processes present a second order
phase transition between a vacuum and a stationary active
state~\cite{Dickman1,Dickman2,Chopard,Schutz,Kwon}. The vacuum
corresponds to an absorbing state towards which the system can
evolve but from which it cannot leave. Systems with such absorbing
states cannot satisfy detailed balance due to the irreversible
character of their dynamics.

Nonequilibrium phase transitions from active fluctuating into
inactive absorbing states have been studied extensively for the
last twenty
years~\cite{Dickman2,Wijland,Dickman3,Fulco1,Lipowski,Albano}.
These phase transitions into absorbing states are characterized by
a nonequilibrium critical behavior similar to that of an
equilibrium phase transition in many respects. One usually uses
the concept of scale invariance to understand nonequilibrium phase
transitions with critical exponents characterizing its universal
behavior. These exponents allow us to categorize nonequilibrium
phase transitions into different universality classes. In reaction
models for the spread of epidemics, the density of infected
individuals $\rho_I$ acts as the order parameter, vanishing in the
absorbing state and having a finite average value in the active
state. The transition between the active and stationary state is
achieved by tuning a typical reaction rate $\lambda$ that acts as
the relevant control parameter.  In the vicinity of the critical
point, the density of infected individuals vanishes as
$\rho\propto (\lambda-\lambda_c)^{\beta}$.  Nonequilibrium phase
transitions are often characterized by spatial and temporal
correlation lengths.  Close to the transition point, the spatial
and temporal correlation lengths diverge as $\xi_s\propto
(\lambda-\lambda_c)^{-\nu}$ and $\xi_t\propto
(\lambda-\lambda_c)^{-\nu '}$.  These two correlation length
scales are related by $\xi_s=\xi_t^z$, where $z=\nu/\nu '$ is
called the dynamic exponent.  These three critical exponents (
$\beta$, $\nu$, $z$) form the basic set of exponents
characterizing the universality class of a given
reaction-diffusion process. It is generally believed that
nonequilibrium phase transitions into absorbing states can be
categorized into a finite number of universality classes. Usually,
systems presenting a continuous transition to a single absorbing
state belong to the directed percolation universality
class~\cite{Dickman2,Hinrichsen1,Lipowski2}. However, some
exceptions to this general rule have been recently reported to
occur basically in systems with a infinite number of absorbing
states~\cite{Jensen,Munoz,Marques,Dickman4,Hooyberghs,Wijland1} or
with particle diffusion~\cite{Fulco2,Hinrichsen2,Kwangho}.

 In recent years, there has been a growing interest in studying diseases caused
by organisms that can mutate into drug resistant forms. One of the
major reasons for the emergence of such resistant organisms is the
inappropriate use of the drugs used to combat them.  Such is the
case, for example, when antibiotics are incorrectly administered
to treat the tuberculosis bacillus~\cite{Gupta,daSilva}, or the
excessive use of anti-malaria drugs in Amazonia,
Brazil~\cite{Ferraroni,Kremsner,deAlencar,Cerutti,Duarte}.
Although these drugs are the only known effective treatments for
such diseases, their abuse can very quickly lead to the emergence
of drug resistant mutant strains. For example, while the exact
mechanisms are not fully understood, strains have evolved that are
resistant to every drug used in Amazonia during the 1950's and
60's. To retard the emergence of such resistant strains, measures
are now being taken in the medical profession to administer drugs
only when needed, and giving preference to drug ``cocktails'' with
multiple components~\cite{deAlencar,Cerutti,Duarte}. Recently, an
epidemic spreading model with immunization and mutations has been
introduced\cite{dammer}. The model also exhibits a phase
transition belonging to the directed percolation universality
class and shows explicitly that the protection gained by
immunization is substantially decreased by the occurrence of
mutations.

In this work, we will study a $1$-dimensional stochastic epidemic
propagation model considering the potential emergence of a
population of drug resistant organisms. We will be mainly
interested in determining the stable phases and the scaling
exponents governing the possible phase transitions. The model
presents two coupled order parameters, namely, the density of drug
sensitive and drug insensitive individuals. The coupling of an
order parameter to a fluctuating density has been shown to modify
the universality class of the absorbing sate phase transition in
some models of population dynamics\cite{Kree,Wijland}. By
employing extensive numerical simulations in finite populations
and finite size scaling analysis, we will show that an active
phase emerges in which there is no drug sensitive individuals but
a finite density of drug resistant ones. The relative strength of
the parameters controlling the processes of healing, death and
development of drug resistance, actually determines the stationary
population state.  Further, we will show that all identified
transitions in this model seem to belong to the directed
percolation universality class.

\section{Basic model for the emergence of drug resistance}
Here, we will follow closely the ideas of \cite{Schinazi}.
Individuals distributed at random on a given environment can be in
three possible states, namely healthy, infected by normal strain and
infected by resistant strain.  Birth of individuals will be introduced
by considering that a vacant site becomes occupied by a healthy
individual at a rate $\beta_1$.  Infection occurs by contact at a rate
$\beta_in_i$ where $n_i$ is the number of infected neighbors and the
index $i$ stands for normal ($i=2$) and drug resistant ($i=3$)
states. Individuals infected with the normal strain can heal at a rate
$p_c$. In what follows, we will consider that the healing rate is due
to drug administration, i.e., the spontaneous healing rate is
negligible. Infected individuals can also develop drug resistance at a
rate $p_r$. This process can be originated by both natural mutations
and bad drug administration. Individuals that develop drug resistance
will not be allowed to heal and die at a mortality rate $p_m$.  The
average surviving-time of drug resistant individuals is much smaller
than the average life-time of individuals infected with the normal
strain, which will be taken as infinite in our simulations.  The above
dynamical rules describing the processes of birth, death, healing,
development of drug resistance and infection by contact with normal or
resistant strains are summarized in table 1.  Notice that the state
with only healthy individuals is an absorbing one, since it is no
longer possible for anybody to become infected. Such is the case, for
example, with small pox, which has been eradicated.

A mean field approximation can be described using the following rate
equations for the average densities:
\begin{eqnarray}
\frac{\partial \rho_1}{\partial t} &=& \beta_1\rho_0 -
\beta_2\rho_1\rho_2 -\beta_3\rho_1\rho_3 + p_c\rho_2 \nonumber \\
\frac{\partial \rho_2}{\partial t} &=& \beta_2\rho_1\rho_2
-p_r\rho_2 -p_c\rho_2 \nonumber \\ \frac{\partial \rho_3}{\partial
t} &=& \beta_3\rho_1\rho_3+p_r\rho_2-p_m\rho_3 \nonumber \\ 1 &=&
\sum_0^3 \rho_i \nonumber
\end{eqnarray}
where a homogeneous, i.e., spatially uniform density of all
populations was assumed. It is straightforward to show that  there
are three possible stationary states: (i)~the absorbing state with
only healthy individuals; (ii)~an active state where healthy
individuals coexist with individuals infected by normal and
resistant organisms; and (iii) a second active state where healthy
and infected individuals coexist but all infected individuals have
the resistant organism. An important point to note is that normal
organisms always coexists with resistant organisms, i.e., a
stationary active state without a drug resistant population is not
possible.

The basic contact process (CP), introduced by Harris~\cite{Harris}
as a ``toy model''  for spreading of disease, corresponds to the
limiting case where only states $1$ and $2$ are
allowed~\cite{Jensen,Munoz,Alon,Dickman5}. The infection reaction
rate is $\lambda n$ where $n$ is the number of infected neighbors,
and the cure rate is $1$, i.e. all sick individuals will recover
in the next time step. Below the critical threshold
$\lambda<\lambda_c$, the absorbing state is reached, but for
$\lambda>\lambda_c$ a stationary active state is possible. The
model can also be applied to study the spread of a disease in the
absence of mutations. In this case, we have only states $0$, $1$
and $2$. This model is known in the literature as the forest fire
model~\cite{Bak,Drossel,Henley}. The above two limits have been
extensively investigated in the literature and accurate estimates
of their critical exponents have shown that they belong to the
directed percolation universality class.  Here, we are going to
perform extensive numerical simulations to obtain the full
phase-diagram of the model allowing all possible states.

\section{Numerical simulations: Phase diagrams and critical exponents}
We employed the above described model in finite one-dimensional chains
with periodic boundary conditions.  Sites are considered to be vacant
or occupied by an individual that interacts with its nearest
neighbors. The initial state corresponds to a random distribution of
healthy ($\eta_1$) and infected ($\eta_2$) individuals.  For a given
set of model parameters, we let the system evolve towards a
statistically stationary state. In what follows we are going to
restrict our study to the particular case of
$\beta_1=\beta_2=\beta_3=1$, letting as free parameters the rates of
mutation ($p_r$), cure ($p_c$) and mortality ($p_m$).  In our
simulations, the statistically stationary state was considered as
reached after $N=10^4$ lattice sweeps. Configurational averages were
performed by considering the subsequent $2\times10^4$ configurations.
Finite size simulations of dynamical systems exhibiting an absorbing
state are usually compromised by the fact that this is the only
stationary state. To avoid this feature, we reactivate the dynamics
whenever the absorbing state is reached by introducing a small density
of the critical order parameter. The dynamics follows the rules of
contamination, healing, mutation, death and birth processes summarized
in table 1. In addition to these rules, the contamination of a healthy
individual that is surrounded by both kinds of infected sites is
chosen to be preferentially updated to state $\eta_3$ (resistant
strain).

In figure~1, we show the equilibrium density of normally infected
individuals ($\rho_2$) and drug resistant ones ($\rho_3$) as obtained
from simulations on chains with $L=800$ sites for three characteristic
values of the mutation rate $p_r$. We observe that the absorbing state
is the equilibrium state for large values of cure and mortality rates.
Two distinct active stationary regimes are also clearly
identified. The first one corresponds to a state with zero density of
normally infected but with a finite density of drug resistant
ones. This phase is stable at low death rates and its emergence is
independent of the mutation rate $p_r\neq 0$. The large surviving time
of drug resistant individuals in this regime favors the contamination
process. The second active state has co-existing densities of normal
and drug resistant infected individuals. It emerges in the regime of
low cure rates but disappears for large mutation rates. A direct
transition between these active states takes place at low death and
cure rates.

In order to closely investigate the transition between the active
state with co-existing densities and the absorbing state, we
computed the order parameter densities for $p_c=0$ (no healing),
which is the most favorable case for co-existence. In figure 2, we
show the relevant densities as function of $p_r$ and $p_m$. Notice
that there is a critical mutation rate above which the phase with
co-existing densities disappears. The maximum value for the
critical mutation rate is attained for $p_m=1$. To have an
accurate estimate of the critical parameters, we employed a finite
size scaling analysis of our numerical data.  Figure~3 shows the
density $\rho_2$ as a function of the mutation rate for the
special case of $p_c=0$ and $p_m=1$ and several chain sizes. The
data suggest a continuous transition near $p_r=0.175$. A numerical
renormalization group analysis of the above data can provide a
better estimate of the critical value. This procedure explores the
finite size scaling relation that the order parameter density is
expected to obey at the vicinity of the critical point, namely
\begin{equation}
\rho_2(L,p_r) = L^{-\beta/\nu}f[(p_r-p_r^*)L^{1/\nu}]\;\;.
\end{equation}
Therefore, one can construct a family of auxiliary functions
\begin{equation}
g(p_r,L,L') = \frac{\ln{\rho_2(L,p_r)/\rho_2(L',p_r)}}{\ln{L/L'}}~~~.
\end{equation}
According to the finite size scaling hypothesis, these auxiliary
functions shall all intercept at the critical point and
$g(p_r^*)=-\beta/\nu$. In figure~4, we report the above scaling
analysis to the data shown in figure~3. All curves obtained from
distinct pairs of lattice sizes intercept nearly at the same
point. Without considering any correction to scaling, we obtain an
accurate estimate of $p_r^*=0.176(1)$ and the critical exponents
ratio $\beta/\nu = 0.25(1)$.  The above finite-size scaling
analysis was employed to obtain the full phase diagram in the
absence of treatment ($p_c=0$) as shown in figure~5. The absorbing
healthy state (H) is always the equilibrium one at large mutation
and mortality rates. The phase diagram exhibits three distinct
phase transitions. For mortality rates above $p_m^* = 0.224(1)$,
the active state is the one with co-existing densities of normal
(N) and resistant (R) individuals which is the stationary state at
small $p_r$. For mortality rates below $p_m^*$ the stationary
state is always active, as the long surviving-time of infected
individuals makes the contamination process the predominant one.
However, there is a transition between the active co-existing
state at small $p_r$ to the active state with a finite density of
drug resistant individuals and a zero density of normal
individuals at large $p_r$.  Notice that at intermediate mutation
rates $p_r$ the phase-diagram presents re-entrant transitions with
the sequence of active-inactive-active equilibrium states
appearing with increasing mortality rates $p_m$.  In all cases,
the measured values of $\beta/\nu$ were consistent with these
transitions belonging to the directed percolation universality
class.  In figure~6 we illustrate the main phase-diagram for a
fixed mutation rate. In this case, the absorbing state is the
equilibrium state at large cure and mortality rates. At low
mortality rates [below the critical value $p_m^*=0.224(1)]$, there
is always a finite density of drug resistant individuals.  With
increasing cure rate $p_c$ a transition between the active
co-existing state and the absorbing state can takes place. The
line delimiting this transition saturates at a maximum critical
cure rate $p_c^*(p_r=0.05) = 0.219(1)$ which decreases with
increasing mutation rates. Further, this co-existing phase is only
stable for mortality rates larger than $p_m^*(p_r=0.05) =
0.168(1)$ and moves to larger values as $p_r$ increases.  At small
mutation rates the above two lines intercept (as shown in the
figure) and a direct transition between the two possible active
states can take place. On the other hand, at the limiting value of
$p_r^*=0.176(1)$ the NR to H transition line disappears and the
co-existing phase is no longer stable. Re-entrant sequence of
active-inactive-active equilibrium states can also be achieved at
intermediate cure rates.
\section{Summary and Conclusions}
We have studied a stochastic epidemic model in a 1-dimensional
lattice which considers the possibility of individuals to develop
drug resistance. Individuals with the normal disease may heal due
to drug administration but those with the resistant strain do not
heal and die at a finite mortality rate. We numerically simulated
this model of population dynamics in finite chains with periodic
boundary conditions and investigated the properties of the
stationary states. The model presents one absorbing inactive state
with only healthy individuals. There is also a main active
stationary state with fluctuating densities of normal and drug
resistant infected individuals. Further, an intermediate active
stationary state with no normally infected individuals and a
fluctuating density of drug resistant ones is also realizable.
This state can be the stationary one only when the mortality rate
is below a critical threshold $p_m^*=0.224$, since under this
condition the long surviving-time of drug resistant individuals
potentialize this epidemic process. For mortality rates above the
critical threshold a transition between the main active to the
absorbing state takes place with increasing cure rates. We
reported typical phase-diagrams showing the stationary states as a
function of the main transition rates. The phase-diagram exhibits
re-entrant regions and a direct transition between the main and
intermediate active states at low mutation rates. We further
employed a finite-size scaling analysis to compute the exponent
ratio $\beta/\nu$ which is consistent with these transitions
belonging to the directed percolation universality class.
\section{Acknowledgments}
This work was partially supported by the Brazilian research agencies
CNPq and CAPES and by the Alagoas state research agency FAPEAL.



\newpage

\begin{table}
\begin{tabular}{|c|c|c|}
\hline ~~~~~~~~~~EVENT~~~~~~~~~
& ~~~~~~~~~~~~~~EFFECT~~~~~~~~~~~~~ & TRANSITION RATE \\
\hline
Birth & $(\eta_0,\eta_1) \rightarrow (\eta_0-1,\eta_1+1)$
& $\beta_1$ \\ Infected by contact & $(\eta_1,\eta_2) \rightarrow
(\eta_1-1,\eta_2+1)$ & $\beta_2n_2$ \\ Infected by contact &
$(\eta_1,\eta_3) \rightarrow (\eta_1-1,\eta_3+1)$ & $\beta_3n_3$ \\
Mutation & $(\eta_2,\eta_3) \rightarrow (\eta_2-1,\eta_3+1)$ & $p_r$
\\ Cure & $(\eta_2,\eta_1) \rightarrow (\eta_2-1,\eta_1+1)$ & $p_c$ \\
Death & $(\eta_3,\eta_0) \rightarrow (\eta_3-1,\eta_0+1)$ & $p_m$ \\
\hline
\end{tabular}
\caption{Summary of the relevant processes, their effects on the
population and transition rates for the present epidemic model with
drug resistance.}
\end{table}

\newpage

\begin{figure}
\centerline{\psfig{figure=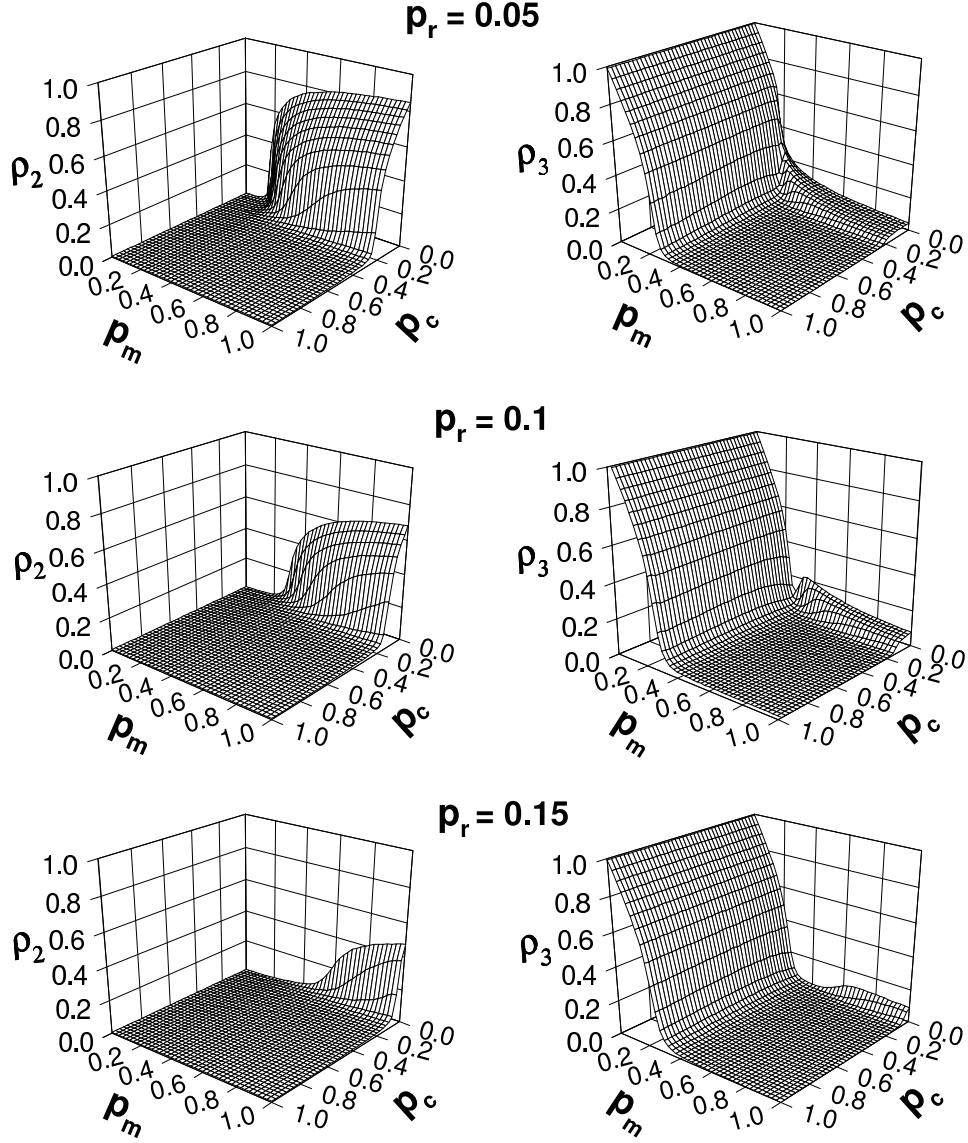,width=5in}}
\medskip
\caption{The stationary densities of normally infected ($\rho_2$) and
drug resistant ($\rho_3$) individuals as a function of cure rate ($p_c$) and
mortality rate ($p_m$) for typical mutation rates.  The three possible
stationary states are clearly identified, namely, the absorbing state
($\rho_2=\rho_3=0$), the main active state (finite $\rho_2$ and
$\rho_3$) and the intermediate active state ($\rho_2=0$ and finite
$\rho_3$). The intermediate active state is practically independent of
the cure and mutation rates. The main active state is only stable for
small mutation rates and disappears for $p_r>p_r^*=0.176$.}
\label{fig1}
\end{figure}

\begin{figure}
\centerline{\psfig{figure=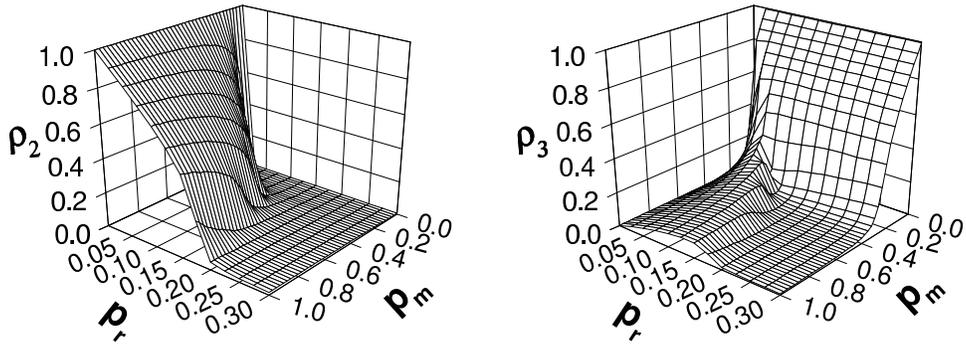,width=5in}}
\caption{The stationary densities of normally infected ($\rho_2$)
and drug resistant ($\rho_3$) individuals as a function of
mutation and mortality rates for the special case of $p_c=0$ (no
healing). Here the transition from the main active to the
absorbing state is better illustrated. In this phase a small
density of drug resistant individuals co-exists with a large
density of normally infected ones.} \label{fig2}
\end{figure}

\clearpage

\begin{figure}
\centerline{\psfig{figure=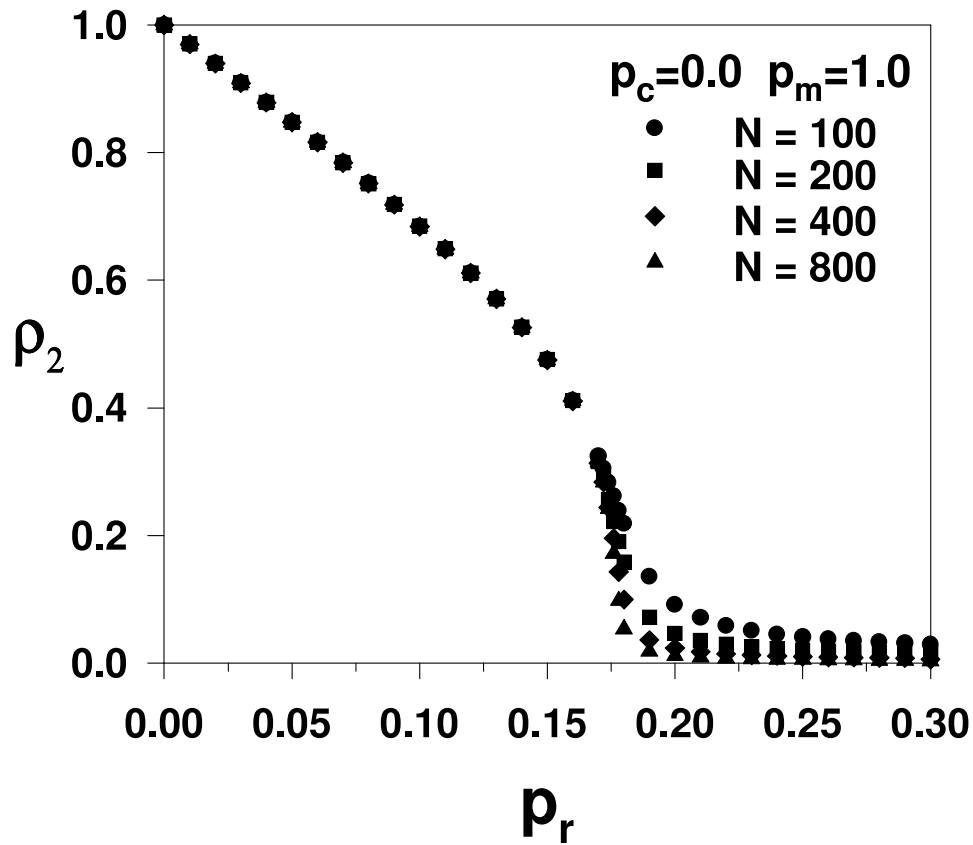,width=5in}}
\caption{ The stationary density of normally infected
individuals ($\rho_2$) versus the mutation rate $p_r$ for the limiting
case of $p_c=0$ and $p_m=1$. Results from simulations on chains with
different sizes are shown. The continuous phase transition from the
active to the absorbing state is smoothed by finite size effects.}
\label{fig3}
\end{figure}

\begin{figure}
\centerline{\psfig{figure=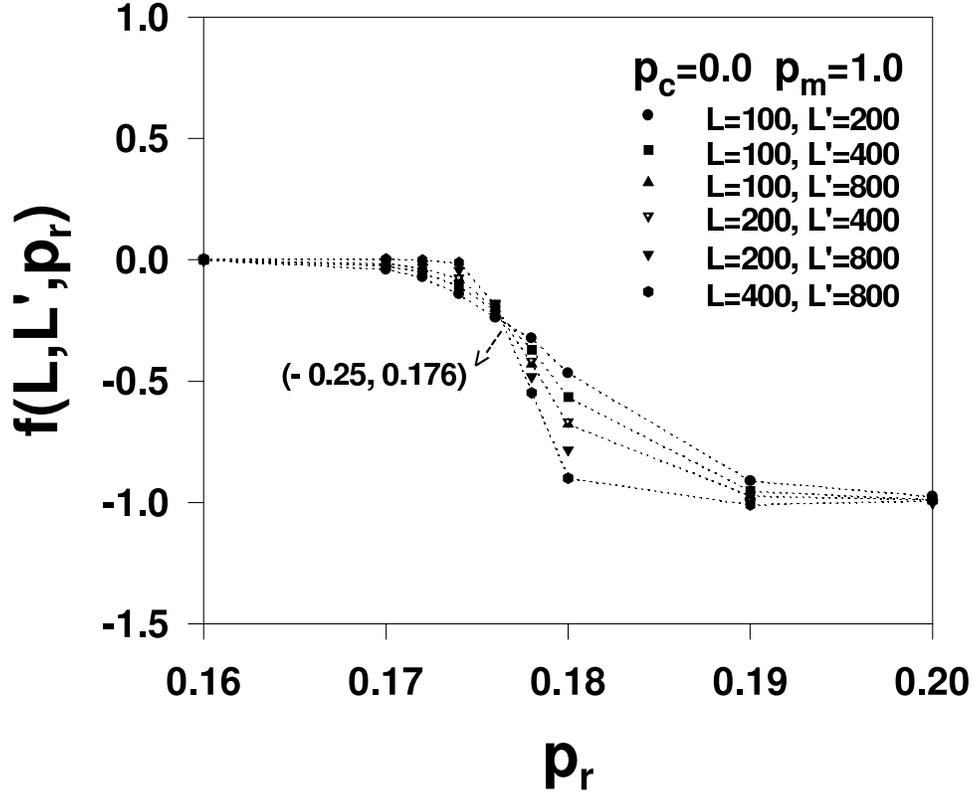,width=5in}}
\caption{ Numerical renormalization group of the same data from
figure~3 in the vicinity of the transition point. Renormalization
using several pairs of chain sizes are shown. All auxiliary
functions intercept at a common point which determines the
critical value $p_r^*=0.176(1)$ and the critical exponents ratio
$\beta/\nu = 0.25(1)$.  This exponent is consistent with this
transition belonging to the directed percolation universality
class.} \label{fig4}
\end{figure}

\begin{figure}
\centerline{\psfig{figure=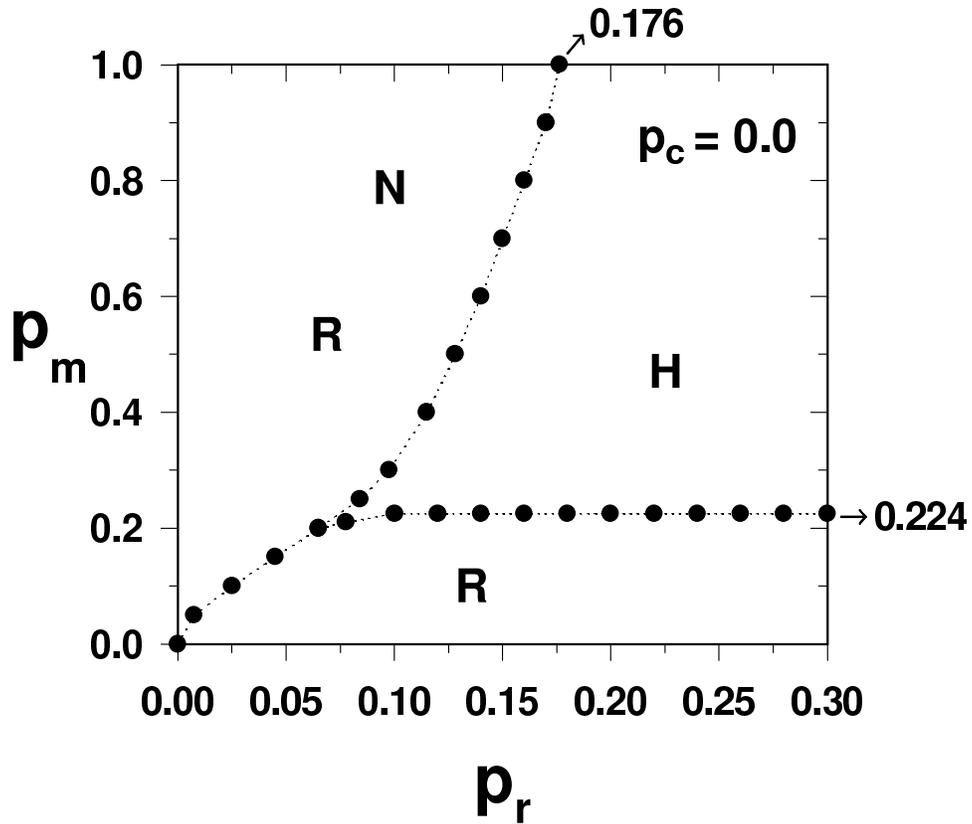,width=5in}}
\caption {Phase-diagram for the special case of $p_c=0$. The
absorbing state (H) is only stable for mortality rates above
$p_m^*=0.224(1)$. In this region, a transition from the main active
(NR) to the absorbing state (H) occurs for increasing mutation rates.
The absorbing state is reached trough the death process. Below $p_m^*$
there is a transition between the main active (NR) to the intermediate
active state (R).  }
\label{fig5}
\end{figure}

\begin{figure}
\centerline{\epsfig{figure=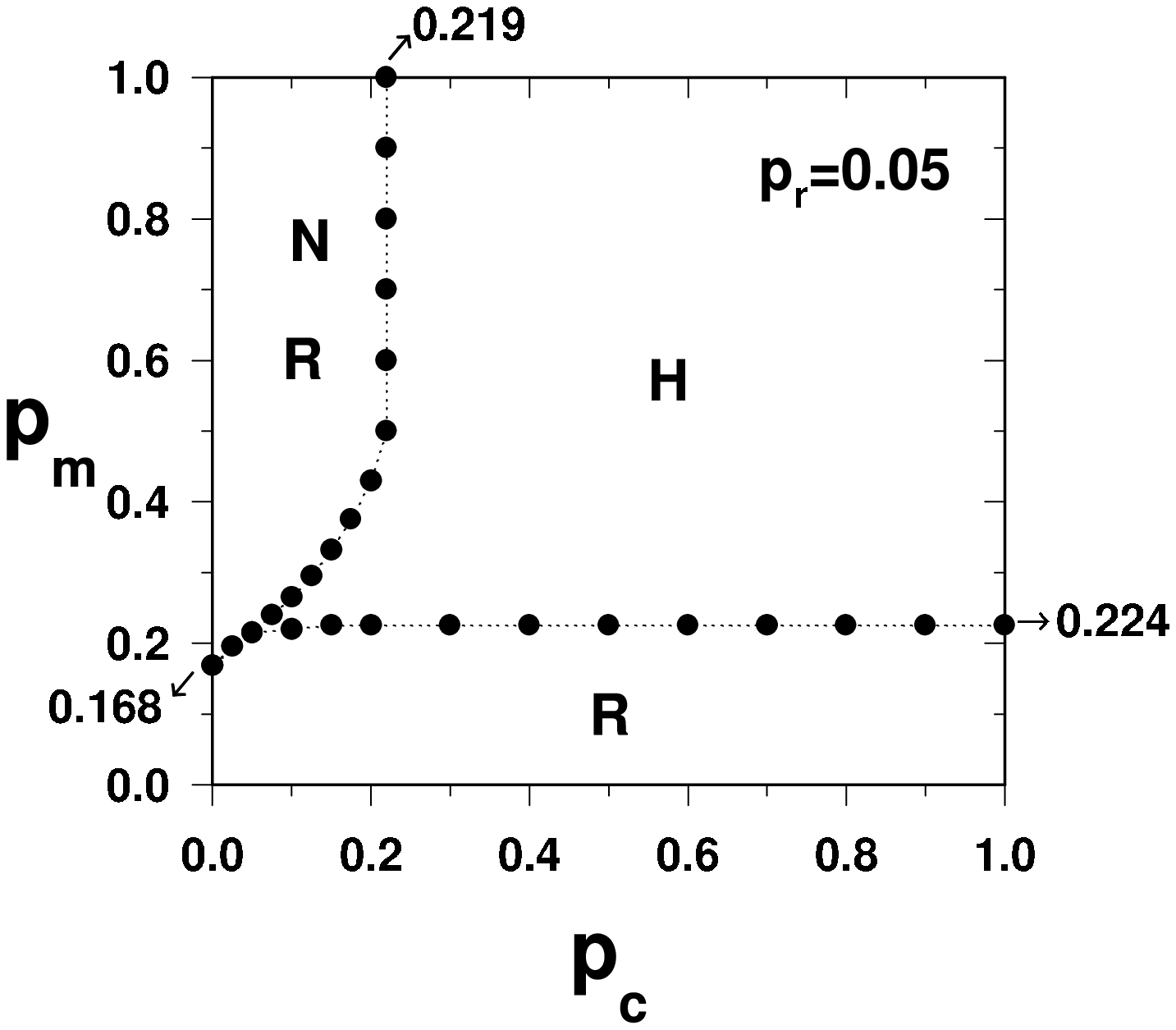,width=5in}}
\caption{ Phase-diagram for a typical
mutation rate $p_r=0.05$.  For low mortality rates the stationary
state is again always active and mostly with drug resistant
individuals.  The main active state may be stable at small cure
rates. Above the critical mortality rate the transition from the
active (NR) to the absorbing (H) state takes place with increasing
cure rates. In this region the absorbing state is reached through both
death and cure process.}
\label{fig6}
\end{figure}


\begin{thebibliography}{mt1}

\bibitem{book} J. D. Murray, {\em Mathematical Biology} (Springer,
Berlin, 1993).
\bibitem{Dickman1} R. Dickman, {\em Nonequilibrium Statistical
Mechanics in One Dimension}, edited by V. Privman (Cambridge
University Press, Cambridge, England,1996).
\bibitem{Dickman2} J. Marro and R. Dickman, {\em Nonequilibrium Phase Transitions
in Lattice Models}, (Cambridge University Press, Cambridge,
England,1999), and references therein.
\bibitem{Chopard} B. Chopard and M. Droz, {\em Cellular Automata
Modelling of Physical Systems}, (Cambridge University Press,
Cambridge, England,1998).
\bibitem{Schutz} G. M. Schutz, {\em Phase Transitions and Critical
Phenomena}, edited by C.Domb and J. L. Lebowitz (Academic Press, New
York, 2000), Vol. 19.
\bibitem{Kwon} S. Kwon, J. Lee and H. Park, Phys. Rev. Lett.
{\bf 85}, 1682 (2000).
\bibitem{Dickman3} R. Dickman,  Physica A {\bf 306}, 90 (2002).
\bibitem{Fulco1} U. L. Fulco, D. N. Messias and M. L. Lyra,
Physica A {\bf 295}, 49 (2001).
\bibitem{Lipowski} A. Lipowski and M. Droz,  Phys. Rev. E {\bf
64}, 031107 (2001).
\bibitem{Albano} E. V. Albano, M. A. Munoz, Phys. Rev. E {\bf
63}, 031104 (2001).
\bibitem{Hinrichsen1} H. Hinrichsen,  Adv. Phys. {\bf 49}, 815
(2000).
\bibitem{Lipowski2} A. Lipowski and D. Lipowska, Physica A {\bf
276}, 456 (2000).
\bibitem{Jensen} I. Jensen, Phys. Rev. Lett. {\bf 70}, 1465
(1993); I. Jensen and R. Dickman,  Phys. Rev. E {\bf 48}, 1710
(1993).
\bibitem{Munoz} M. A. Mu\~noz, G. Grinstein, R. Dickman and R. Livi,
Phys. Rev. Lett.  {\bf 76}, 451 (1996).
\bibitem{Marques} M. C. Marques, M. A. Santos and J. F. F. Mendes,
Phys. Rev. E {\bf 65}, 016111 (2001).
\bibitem{Dickman4} R. Dickman, W. R. M. Rab\^elo and G. \'Odor, Phys.
Rev. E {\bf 65}, 016118 (2001).
\bibitem{Hooyberghs} J. Hooyberghs, E. Carlon and C. Vanderzande,
Phys. Rev. E {\bf 64}, 036124 (2001).
\bibitem{Wijland1} F. van Wijland, Phys. Rev. Lett. {\bf 89}, 190602
(2002).
\bibitem{Fulco2} U. L. Fulco, D. N. Messias and M. L. Lyra, Phys.
Rev. E {\bf 63}, 066118 (2001).
\bibitem{Hinrichsen2} H. Hinrichsen,  Phys. Rev. E {\bf 63},
036102 (2001).
\bibitem{Kwangho} Kwangho Park and In-mook Kim,  Phys. Rev. E
{\bf 66}, 027106 (2002).
\bibitem{Gupta} R. Gupta, J. G. Brenner, C. L. Henry, J. Y. Kim, S. Shin,
M. Espinal, M. C. Raviglione,  WHO {\bf 276}, 1-20 (1999), and
references therein.
\bibitem{daSilva} P. E. A. da Silva, M. Os\'orio, M. C. Reinhardt, L. S.
Fonseca and O. A. Dellagostin,  Microbes and Infection {\bf 3},
1111 (2001).
\bibitem{Ferraroni} J. J. Ferraroni, F. H. Alencar and R. Shrimpton,
 T. Roy. Soc.  Trop. Med. Hyg. {\bf 77}, 138, (1983); J. J.
Ferraroni, Rev. Saude Publ. {\bf 17}, 328, (1983); F. H. Alencar,
J. J. Ferraroni and R. Shrimpton,  Rev. Saude Publ. {\bf 16}, 299,
(1982).
\bibitem{Kremsner} P. G. Kremsner, G. M. Zotter, W. Grainger and H.
Feldmeier,  Lancet {\bf 2}, 684, (1987); P. G. Kremsner, H.
Feldmeier, R. M. Rocha and W.  Grainger, Rev. I. Med. Trop. {\bf
30}, 118, (1988)
\bibitem{deAlencar} F. E. C. de Alencar, C. Cerutti, R. R. Durlacher,
M. Boulos, F. P. Alves, W. Milhous and L. W. Pang,  J. Infect Dis.
{\bf 175}, 1544, (1997).
\bibitem{Cerutti} C. Cerutti, R. R. Durlacher, F. E. C. de Alencar, A.
A. C Segurado and L. W. Pang, J. Infect Dis. {\bf 180}, 2077,
(1999).
\bibitem{Duarte} E. C. Duarte, L. W. Pang, L. C. Ribeiro and C. J. F.
Fontes,  Am. J. Trop. Med. Hyg. {\bf 65}, 471, (2001).
\bibitem{dammer} S. M. Dammer and H. Hinrichsen, Phys. Rev. E {\bf
68}, 016114 (2003).
\bibitem{Kree} R. Kree, B. Schaub, and B. Schmittmann, Phys. Rev. A {\bf 39}, 2214
(1989).
\bibitem{Wijland} K. Oerding, F. van Wijland, J.P. Leroy,
and H.J. Hilhorst, J. Stat. Phys. {\bf 99}, 1365 (2000).
\bibitem{Schinazi} R. B. Schinazi,  J. Stat. Phys. {\bf 97}, 409,
(1999).
\bibitem{Harris} T. E. Harris, Ann. Prob. {\bf 2}, 969 (1974).
\bibitem{Alon} U. Alon, M. R. Evans, H. Hinrichsen and D. Mukamel,
Phys. Rev. Lett. {\bf 76}, 2746 (1996).
\bibitem{Dickman5} R. Dickman, J. K. L. da Silva,  Phys. Rev. E
{\bf 58}, 4266 (1998).
\bibitem{Bak} P. Bak, K. Chen, and C. Tang,  Phys. Lett. A {\bf
147}, 297 (1990).
\bibitem{Drossel} B. Drossel and F. Schwabl,  Phys. Rev. Lett. {\bf
69}, 1629 (1992); for a review, see S. Clar, B. Drossel, and F.
Schwabl, J.  Phys.: Condens. Matter {\bf 8}, 6803 (1996).
\bibitem{Henley} C. L. Henley,  Bull. Am. Phys. Soc. {\bf 34}, 838
(1989);  Phys. Rev. Lett. {\bf 71}, 2741 (1993).
\end{thebibliography}
\end{document}